\documentclass[sigconf]{acmart}

\AtBeginDocument{%
  }

\newcommand{\ytm}{YouTube Music\xspace}
\usepackage{xspace}
\usepackage{graphicx}

\copyrightyear{2026}
\acmYear{2026}
\setcopyright{cc}
\setcctype{by}
\acmConference[RecSys '26]{20th ACM Conference on Recommender Systems}{September 27-October 02, 2026}{Minneapolis, MN, USA}
\acmBooktitle{20th ACM Conference on Recommender Systems (RecSys '26), September 27-October 02, 2026, Minneapolis, MN, USA}
\acmDOI{10.1145/3773078.3831876}
\acmISBN{979-8-4007-2284-4/2026/09}

\begin{document}

\title{Breaking the Loop: An Empirical Comparison of Strategies for Novelty and Freshness in YouTube Music}

\author{Srivaths Ranganathan}
\email{srivaths@google.com}
\affiliation{%
  \institution{Google LLC}
  \city{Mountain View}
  \country{USA}
}

\author{Zihuan Diao}
\email{zdiao@google.com}
\affiliation{%
  \institution{Google LLC}
  \city{Mountain View}
  \country{USA}
}

\author{Bernardo Cunha}
\email{becunha@google.com}
\affiliation{%
  \institution{Google LLC}
  \city{New York}
  \country{USA}
}

\author{Joshua L. Moore}
\email{jlmo@google.com}
\affiliation{%
  \institution{Google LLC}
  \city{New York}
  \country{USA}
}

\author{Robin Dumas}
\email{dumasrobin@google.com}
\affiliation{%
  \institution{Google LLC}
  \city{Mountain View}
  \country{USA}
}

\author{Murat Goksedef}
\email{mgoksedef@google.com}
\affiliation{%
  \institution{Google LLC}
  \city{Mountain View}
  \country{USA}
}
\author{Yanwei Song}
\email{yanweisong@google.com}
\affiliation{%
  \institution{Google LLC}
  \city{Mountain View}
  \country{USA}
}
\author{Mukai Lu}
\email{mukail@google.com}
\affiliation{%
  \institution{Google LLC}
  \city{Mountain View}
  \country{USA}
}

\author{Gergo Varady}
\email{gvarady@google.com}
\affiliation{%
  \institution{Google LLC}
  \city{New York}
  \country{USA}
}

\author{Tracy Pesin}
\email{tracypesin@google.com}
\affiliation{%
  \institution{Google LLC}
  \city{New York}
  \country{USA}
}

\renewcommand{\shortauthors}{Ranganathan et al.}

\begin{abstract}
Continuously trained ranking models in music recommenders fall into feedback loops where previously consumed items dominate recommendations. This suppresses two distinct content classes: new releases (temporal freshness) and unlistened catalog items (novelty). Industry practitioners have a wide menu of interventions available, ranging from serving-time heuristics, training-data reweighting, architectural debiasing, to uncertainty-driven exploration, each of which are well understood in academic settings. But live systems offer challenges with continuously ingested content, interconnected components, and practical limitations that counteract the findings from academic research. 

We report results from off-policy online A/B tests for six interventions and a combination experiment across four conceptual layers (serving, training, architecture, exploration) on the YouTube Music homepage. All interventions modify the ranking model or the serving layer that consumes its scores; candidate generation and other upstream components are held fixed. We discuss key takeaways from our results: first, serving-time interventions on continuously trained systems are neutralized by the learning loop. Second, architectural debiasing reduces popularity dominance and improves diversity but does not create discovery, while carrying hidden integration costs. Finally, uncertainty-driven exploration interventions with a Spectral-normalized Neural Gaussian Process (SNGP) head produce the largest new-release lift, though they come with a measurable engagement or diversity tradeoff. We close with recommendations on which layer to intervene at, and the hidden costs of each choice.

\end{abstract}



\maketitle

\section{Introduction}

Music recommendation has an unusual property: listeners return to a small set of familiar tracks (repeat consumption) but also expect the system to surface \emph{new releases}, and to introduce them to \emph{novel} artists and songs they have not heard before. A recommender that perfectly predicts replay fails at the latter two, and over a long enough horizon fails at the first as well, as the user's breadth of content stops growing and engagement plateaus or even declines.

The mechanism that produces this failure is well understood in the literature ~\cite{jiang2019degenerate}. A ranking model trained on logged engagement learns that previously consumed items convert at a high rate. Continuously retraining the model on logged data closes the loop: items that were ranked highly receive impressions, those impressions generate engagement, and the next training run is fit on data that further reinforces the same items. New releases arrive into a system already saturated with engaged-on alternatives; discovery candidates compete against an internal ranking of items the user is already known to like. Beyond the user-facing impact on novelty and freshness, this dynamic also degrades the long-term health of the recommender system itself: a training distribution dominated by repeated interactions provides weak signal for items that have not yet been surfaced, and the system loses the ability to evaluate unseen items.

A wide range of interventions has been proposed to counter this dynamic. They include serving-time score adjustments such as recency boosts, training-data reweighting via inverse propensity scoring (IPS) ~\cite{schnabel2016recommendations} , architectural changes such as the position-bias towers used in PAL ~\cite{guo2019pal} and YouTube's WatchNext multitask ranker ~\cite{zhao2019recommending}, and exploration strategies based on bandits and posterior sampling. Each makes a different assumption about how best to correct the failure. Heuristic boosts treat the symptom at the serving layer. Both IPS-style reweighting and position bias towers assume the failure originates from logged data being biased by what the system previously chose to show; however, they intervene at different layers. IPS addresses this at the data layer by reweighting the biased samples, whereas bias towers treat presentation confounds as architectural artifacts to be jointly modeled during training and removed at serving. Uncertainty-driven exploration treats the training signal itself as too narrow, and uses model uncertainty to direct attention to items the system has not learned about yet. In our experience, these mental models do not all map onto the same failure, and choosing the wrong layer wastes both engineering time and online traffic.\\

\noindent
\textbf{Contributions} 
\noindent
\\
This paper makes three contributions. (1) A \emph{layered} view of feedback-loop interventions, organized by where in the ranker stack the change is applied: serving, training data, model architecture, or exploration policy. (2) Online A/B results for six interventions plus one combination experiment, evaluated against the same control on the \ytm homepage. To our knowledge this is the first head-to-head online comparison of this breadth on a single production music surface. (3) Practitioner-facing claims supported by empirical evidence on the product impact, hidden costs and downstream system interactions associated with intervening at each of these conceptual layers.\\

\noindent
\textbf{Why this is hard to study}
\noindent
\\
A recurring difficulty in industry recommender research is that online metrics from a short A/B do not fully capture the second-order effects of increased exposure for new content. The rest of the system: candidate generators, filters, other interacting scorer components and downstream ranking functions adapt to the shift in impressions in ways that amplify or even nullify the intended effect. On-policy experiments at the scale of large ranking models are difficult to set up cleanly, since the large models can't be trained on a small experimental dataset. This makes it hard to predict in advance how the broader system will respond to changes in the impression distributions of a slice of content. While our study is subject to these practical constraints, we explicitly document these system adaptations in our results. These downstream systemic adaptations are often further compounded by organic, off-platform demand signals, making the isolation of recommendation-driven freshness gains a hard causal inference problem~\cite{mehrotra2020inferring}. We believe highlighting these dynamics is part of why this empirical comparison is useful for practitioners.

\section{Related Work}

\paragraph{Cold start and new content.} The cold-start problem~\cite{schein2002methods} is the classical formulation of the new-content failure mode: items with no interaction history cannot be ranked from collaborative signal alone. Industrial recommenders typically address cold start with content-based features in the candidate generator and ranker; deep models trained on user-item engagement, including the YouTube architecture in~\cite{covington2016deep} and its multi-task successor in~\cite{zhao2019recommending}, have largely absorbed cold-start handling into the feature pipeline. Cold start, however, is a subset of the broader problem we address: a song newly released by an artist a user already follows is not cold in the user-feature sense, but is cold in the system-engagement sense.  Without historical positive engagement to drive up the model's confidence, the model defaults to lower baseline predictions and these items systematically rank below better-trained alternatives.

\paragraph{Popularity and exposure bias.} Popularity bias has been studied extensively from both algorithmic and fairness perspectives~\cite{abdollahpouri2019unfairness}. Anderson et al.~\cite{anderson2020algorithmic} document that algorithmic listening on Spotify is consistently less diverse than user-driven listening, and that diversification on the platform is associated with users shifting away from algorithmic surfaces. We treat popularity bias as a related but distinct problem from new-release suppression: an intervention can reduce popularity dominance without lifting new releases, and we observe exactly this pattern with the architectural debiasing approach we evaluate.

\paragraph{Counterfactual learning and IPS} Schnabel et al.~\cite{schnabel2016recommendations} formalized recommendations-as-treatments and introduced inverse propensity scoring as a tool for unbiased learning and evaluation under selection bias. Joachims et al.~\cite{joachims2017unbiased} extended the approach to learning-to-rank from biased click logs. To address sample selection bias—where logs only reflect shown documents—Ovaisi et al. \cite{ovaisi2020correcting} introduced counterfactual approaches adapting Heckman’s two-stage method across pointwise, pairwise, and listwise ranking objectives. Bonner and Vasile~\cite{bonner2018causal} proposed causal embeddings that further leverage the distinction between organic and recommended interactions. Furthermore, Wang et al. \cite{wang2018position} proposed a listwise cross-entropy framework to estimate position bias without randomized traffic.

\paragraph{Position bias and architectural debiasing.} The PAL framework~\cite{guo2019pal} models position bias as a multiplicative factor in CTR prediction and trains a position-aware module that is dropped at serving. Zhao et al.~\cite{zhao2019recommending} introduce a shallow tower in YouTube's WatchNext multitask ranker that takes presentation features (including position and device) as inputs to produce a bias correction added to the main model's logits, again dropped at serving. Both architectures bake the assumption that the dominant logged-data confound is position and presentation, and both share the property that the resulting unbiased predictor preserves ranking but shifts the distribution of predicted scores. As identified by Zhang et al. \cite{zhang2023towards}, this setup overlooks a causal confounding factor: because display positions in logged data are determined by the previous production model, position features are heavily entangled with true relevance. Without explicit disentanglement, the bias tower leaks relevance signal, meaning that dropping the tower at serving inadvertently discards key preference information. A recent large-scale study by Hager et al. \cite{hager2024unbiased} evaluating pointwise, pairwise, and listwise debiasing on Baidu's search logs demonstrated that offline click-prediction gains do not easily convert into online relevance improvements, highlighting the practical difficulty of deploying these models in production.

\paragraph{Multi-task learning for ranking.} Modern industrial rankers, including the system we deploy on, predict multiple engagement targets jointly~\cite{ zhao2019recommending}. Multi-gate Mixture-of-Experts~\cite{ma2018modeling} is a common building block. The relevance for this paper is that auxiliary heads, whether for position bias, for debiasing, or for uncertainty estimation, are added to a multi-task objective whose loss-weight balance is itself a tunable system. While multi-task learning addresses multiple objectives within a single-model paradigm, alternative decentralized structures such as multi-agent architectures  ~\cite{ranganathan2026multi} have recently been proposed to decompose recommendation subtasks (e.g., separating perception from reasoning and feedback integration) to cooperatively balance competing system-level goals like diversity, fairness, and long-term user value. Some of our findings touch on this: simply adding an auxiliary head to the objective can shift behavior even when the head is not consumed at serving.

\paragraph{Exploration and bandits.} The UCB family~\cite{auer2002finite} provides the textbook foundation for uncertainty-driven exploration. In recommender systems, exploration has been studied in contextual bandit form~\cite{li2010contextual} and via off-policy correction for policy-gradient learners~\cite{chen2019topk}. In production music streaming, active exploration via multi-arm bandits has been deployed to balance this trade-off for new release carousels, though simpler neural predictions of cold-start embeddings can often perform on par due to slow bandit convergence on rapidly evolving fresh content pools~\cite{briand2024let}. We borrow the UCB idea at serving time and the curriculum-learning idea~\cite{bengio2009curriculum} at training time, with per-item uncertainty supplied by a spectral-normalized Gaussian process head~\cite{liu2020simple}. Furthermore, how users explore newly discovered music clusters is closely coupled with their intrinsic discovery needs and follows structured trajectories; listeners often initiate exploration into unfamiliar genres with highly popular, accessible tracks before moving deeper into niche, representative items~\cite{moscati2025familiarizing}.

\paragraph{Feedback loops.} Chaney et al.~\cite{chaney2018algorithmic} use simulation to show that algorithmic confounding homogenizes user behavior without corresponding utility gains. Jiang et al.~\cite{jiang2019degenerate} characterize degenerate feedback loops formally. Mansoury et al.~\cite{mansoury2020feedback} document feedback-driven amplification of popularity bias in offline simulation. In music recommendation contexts, these feedback loops are often exacerbated by the fact that highly optimized, satisfaction-centric ranking models naturally exhibit an increased bias toward popular and historically familiar content as model complexity grows~\cite{hansen2021shifting}. Our contribution complements this line of work with online evidence on a production music surface, and a head-to-head comparison of interventions that take different views of where the loop should be broken.

\section{Problem Description}\label{sec:desc}

We study the homepage surface of the \ytm app. Items eligible for ranking on this surface include songs, albums, playlists, artists and algorithmic mixes. The production ranker is a deep multi-task model in the lineage of~\cite{covington2016deep, zhao2019recommending}, predicting click-through rate, watch time, and several other engagement targets. The model is trained continuously on the past 4 weeks of logged impressions and engagements, and a new ranker is shipped on a regular cadence. The model also distills predictions \cite{ranganathan2025zero} from a larger teacher model trained on YouTube's data. We use the term \emph{new release} for songs and albums released within 7 days of the impression timestamp. We use \emph{discovery} for items the user has had no prior engagement with on the platform; discovery and new-release sets overlap but are not equal, and an item can be a new release for an artist the user is already familiar with. The product goal of this study is to lift exposure for both classes without unacceptably degrading top-line homepage engagement.\\

\noindent
\textbf{Debiasing is not a freshness intervention} \\
\noindent
As we outline our framework, we make an explicit distinction between debiasing and freshness. Practitioners often reach for popularity biasing tools when the product problem is to surface new content. Debiasing is a mathematical correction: it aims to correct a ranker that trains on biased logged data (e.g., position or popularity bias). Freshness, on the other hand, is a product outcome: the goal is to surface new content that the user has not seen. New releases lack the history of engagement that would let the ranker score them fairly and debiasing alone can't produce the required signal for these items. An ideally debiased model may still systematically under-rank new releases simply because they lack the historical signal needed to generate a confident positive score, whereas a heuristic boost can lift new releases without correcting any underlying bias. We return to this idea in subsection \ref{subsec:debias_freshness} when discussing the impact of the bias tower on freshness. \\

\section{Layered Framework and Methods}\label{sec:methods}

We evaluate six interventions, organized into four conceptual layers based on where in the system they are applied. All are deployed in the same production ranker described in Section 3 and evaluated against an unmodified baseline. Each intervention is implemented as a minimal change to the smallest reasonable component: we did not stack interventions across layers, and each experimental arm differs from control in a single specified way.

\subsection{Serving}
These are changes that apply only in the ranking layer that consumes the scores from the models and combines it to generate a single final ranked list. Interventions here, such as heuristic recency score boosts, adjust the presentation policy directly without requiring any model retraining.\\

\noindent\textbf{Approach 1 -- Heuristic recency boost at serving.} A multiplicative score boost is applied to items released within $N$ days, applied at the ranker's score-combination step. No model is retrained. This is the lightest possible intervention and the one most teams attempt first; it serves both as a baseline and as a probe of how the surrounding system responds when ranking is perturbed at serving. We experimented with different factors. All arms had a similar behavior for the New release metrics but with larger swings in impressions over days.\\

\noindent\textbf{Approach 2 -- Diversification by release age.} A soft demotion is applied to diversify the items across buckets based on the number of days since the item was released. This diversification applies alongside other diversification factors (e.g. across artists) and adding a new dimension of diversification naturally reduces the impact of existing factors.

\subsection{Training Data Debiasing}
Interventions at this layer modify the distribution or importance of the logged examples fed into the model. Common approaches include temporal sample reweighting (e.g., up-weighting more recent days) or inverse propensity scoring (IPS) or product informed weights adjusting for bias in the presentation on the platform (e.g., for items at the edge of the screen). We did not run a live experiment with IPS as the specific variants we tried ended up being unstable during training. We implemented IPS using the Predicted CTR (PCTR) from the logged predictions with a few variations for stability such as clipping the weights and normalizing by the global median PCTR. \\

\noindent\textbf{Approach 3 -- Up-weight data from recent days.} Examples in the 28-day training window are reweighted with an exponential decay to give higher weight to more recent days. The model architecture and serving stack are unchanged. Hypothesis: Because new releases (by definition less than 7 days old) only possess engagement signals in the most recent days of our 28-day training window, a uniform weighting (across time) drowns out their sparse signals in favor of established items. By applying an exponential decay to up-weight recent days, we disproportionately increase the relative training impact of the narrow time window where new releases actually exist, allowing the model to adapt faster to newly trending content rather than fitting its capacity to older, accumulated engagement patterns.

\begin{table*}[hbtp]
\centering
\caption{Experimental results of various interventions on new releases, discovery, diversity and homepage engagement. Positive statistically significant changes are indicated in \textbf{bold}, and negative statistically significant changes are indicated in \textit{italics}. Arrows ($\uparrow$, $\downarrow$) indicate the direction of the statistically significant change at the 95th percentile confidence intervals.}\label{tab:results}
\resizebox{\textwidth}{!}{
\begin{tabular}{lcccccc}
\toprule
\textbf{Technique} & \textbf{New Releases} & \textbf{New Releases} & \textbf{Novel Music} & \textbf{Artist} & \textbf{Homepage} & \textbf{Homepage} \\  
 & \textbf{1 day Engagement} & \textbf{7 day Engagement} & \textbf{Engagement} & \textbf{Diversity} & \textbf{Engagement} & \textbf{Repetition *} \\
\midrule
1. Heuristic recency boost at serving & \begin{tabular}{@{}c@{}}-0.74\% \\ {[-2.98, 1.50]}\end{tabular} & \begin{tabular}{@{}c@{}}-0.72\% \\ {[-2.36, 0.91]}\end{tabular} & \begin{tabular}{@{}c@{}}-0.26\% \\ {[-0.77, 0.25]}\end{tabular} & \begin{tabular}{@{}c@{}}-0.09\% \\ {[-0.22, 0.04]}\end{tabular} & \begin{tabular}{@{}c@{}}\textit{-0.28\%$^{\downarrow}$} \\ \textit{[-0.46, -0.10]}\end{tabular} & \begin{tabular}{@{}c@{}}\textit{-0.32\%$^{\downarrow}$} \\ \textit{[-0.40, -0.25]}\end{tabular} \\
\addlinespace
2. Diversification by release age & \begin{tabular}{@{}c@{}}1.15\% \\ {[-1.76, 4.07]}\end{tabular} & \begin{tabular}{@{}c@{}}0.38\% \\ {[-1.45, 2.20]}\end{tabular} & \begin{tabular}{@{}c@{}}-0.06\% \\ {[-0.53, 0.42]}\end{tabular} & \begin{tabular}{@{}c@{}}\textit{-1.17\%$^{\downarrow}$} \\ \textit{[-1.57, -0.76]}\end{tabular} & \begin{tabular}{@{}c@{}}0.05\% \\ {[-0.12, 0.23]}\end{tabular} & \begin{tabular}{@{}c@{}}\textbf{0.22\%$^{\uparrow}$} \\ \textbf{[0.13, 0.31]}\end{tabular} \\
\addlinespace
3. Up-weight data from recent days & \begin{tabular}{@{}c@{}}-3.33\% \\ {[-6.94, 0.28]}\end{tabular} & \begin{tabular}{@{}c@{}}-0.87\% \\ {[-2.69, 0.95]}\end{tabular} & \begin{tabular}{@{}c@{}}\textbf{0.91\%$^{\uparrow}$} \\ \textbf{[0.35, 1.46]}\end{tabular} & \begin{tabular}{@{}c@{}}\textit{-0.27\%$^{\downarrow}$} \\ \textit{[-0.44, -0.10]}\end{tabular} & \begin{tabular}{@{}c@{}}0.07\% \\ {[-0.17, 0.32]}\end{tabular} & \begin{tabular}{@{}c@{}}0.00\% \\ {[-0.10, 0.09]}\end{tabular} \\
\addlinespace
4. Non Serving Position Bias tower & \begin{tabular}{@{}c@{}}0.37\% \\ {[-2.33, 3.07]}\end{tabular} & \begin{tabular}{@{}c@{}}\textit{-3.39\%$^{\downarrow}$} \\ \textit{[-4.53, -2.24]}\end{tabular} & \begin{tabular}{@{}c@{}}\textbf{4.70\%$^{\uparrow}$} \\ \textbf{[4.23, 5.17]}\end{tabular} & \begin{tabular}{@{}c@{}}\textbf{0.79\%$^{\uparrow}$} \\ \textbf{[0.64, 0.93]}\end{tabular} & \begin{tabular}{@{}c@{}}-0.04\% \\ {[-0.18, 0.09]}\end{tabular} & \begin{tabular}{@{}c@{}}\textit{-0.53\%$^{\downarrow}$} \\ \textit{[-0.61, -0.46]}\end{tabular} \\
\addlinespace
5 (a). Uncertainty prediction (non serving) & \begin{tabular}{@{}c@{}}\textbf{5.32\%$^{\uparrow}$} \\ \textbf{[2.68, 7.96]}\end{tabular} & \begin{tabular}{@{}c@{}}\textbf{1.87\%$^{\uparrow}$} \\ \textbf{[0.94, 2.80]}\end{tabular} & \begin{tabular}{@{}c@{}}0.29\% \\ {[-0.20, 0.78]}\end{tabular} & \begin{tabular}{@{}c@{}}0.02\% \\ {[-0.38, 0.42]}\end{tabular} & \begin{tabular}{@{}c@{}}-0.05\% \\ {[-0.28, 0.18]}\end{tabular} & \begin{tabular}{@{}c@{}}\textit{-0.11\%$^{\downarrow}$} \\ \textit{[-0.20, -0.03]}\end{tabular} \\
\addlinespace
5 (b). Uncertain items boosted at serving & \begin{tabular}{@{}c@{}}\textbf{7.75\%$^{\uparrow}$} \\ \textbf{[4.81, 10.70]}\end{tabular} & \begin{tabular}{@{}c@{}}\textbf{3.30\%$^{\uparrow}$} \\ \textbf{[2.01, 4.60]}\end{tabular} & \begin{tabular}{@{}c@{}}0.01\% \\ {[-0.49, 0.51]}\end{tabular} & \begin{tabular}{@{}c@{}}\textit{-0.95\%$^{\downarrow}$} \\ \textit{[-1.43, -0.47]}\end{tabular} & \begin{tabular}{@{}c@{}}-0.05\% \\ {[-0.26, 0.17]}\end{tabular} & \begin{tabular}{@{}c@{}}\textit{-0.94\%$^{\downarrow}$} \\ \textit{[-1.02, -0.86]}\end{tabular} \\
\addlinespace
6. Uncertainty-weighted training loss & \begin{tabular}{@{}c@{}}\textbf{4.33\%$^{\uparrow}$} \\ \textbf{[1.70, 6.97]}\end{tabular} & \begin{tabular}{@{}c@{}}0.77\% \\ {[-0.48, 2.01]}\end{tabular} & \begin{tabular}{@{}c@{}}\textit{-1.46\%$^{\downarrow}$} \\ \textit{[-1.91, -1.02]}\end{tabular} & \begin{tabular}{@{}c@{}}\textbf{0.17\%$^{\uparrow}$} \\ \textbf{[0.01, 0.33]}\end{tabular} & \begin{tabular}{@{}c@{}}\textit{-0.40\%$^{\downarrow}$} \\ \textit{[-0.60, -0.20]}\end{tabular} & \begin{tabular}{@{}c@{}}\textit{-0.23\%$^{\downarrow}$} \\ \textit{[-0.30, -0.17]}\end{tabular} \\
\addlinespace
7. Uncertainty-weighted training loss + bias tower$^{\dagger}$ & \begin{tabular}{@{}c@{}}2.00\% \\ {[-1.24, 5.25]}\end{tabular} & \begin{tabular}{@{}c@{}}\textit{-3.41\%$^{\downarrow}$} \\ \textit{[-5.30, -1.52]}\end{tabular} & \begin{tabular}{@{}c@{}}\textbf{2.92\%$^{\uparrow}$} \\ \textbf{[2.36, 3.48]}\end{tabular} & \begin{tabular}{@{}c@{}}\textbf{0.90\%$^{\uparrow}$} \\ \textbf{[0.77, 1.03]}\end{tabular} & \begin{tabular}{@{}c@{}}\textit{-0.43\%$^{\downarrow}$} \\ \textit{[-0.55, -0.30]}\end{tabular} & \begin{tabular}{@{}c@{}}\textit{-0.48\%$^{\downarrow}$} \\ \textit{[-0.58, -0.37]}\end{tabular} \\
\bottomrule
\multicolumn{7}{l}{\footnotesize * For Homepage Repetition, a reduction is the desired outcome.} \\
\multicolumn{7}{l}{\footnotesize $^{\dagger}$ Metrics are from only one week of data for this arm.}
\end{tabular}
}
\end{table*}

\subsection{Architectural Debiasing}

This layer includes additions or changes to the model architecture, typically auxiliary heads or features that allow the model to factor the bias out of its predictions. These modules explicitly model confounding factors (like position or client bias) during training so that the main prediction heads learn a fairer, debiased representation.\\

\noindent\textbf{Approach 4 -- Shallow position/client bias tower (architecture).} A shallow auxiliary tower takes position and client-type sparse features and outputs a CTR correction term that is added to the main model's logits during training; the tower is dropped at serving. The architecture follows PAL~\cite{guo2019pal} and the WatchNext shallow tower~\cite{zhao2019recommending}. Hypothesis: the top ranked items on the Homepage are often previously consumed quick-access familar tracks. By accounting for and removing the bias from the item position, the ranker should produce fairer scores for typically lower ranked items.

\subsection{Exploration}
Interventions that leverage estimates of model uncertainty to actively guide the system toward items it knows less about. This exploration signal can be applied either at serving time (by adding an uncertainty boost to the item’s score) or at training time (by up-weighting high-uncertainty examples in the loss).\\

\noindent\textbf{Approach 5 -- Uncertain items boosted at serving.} A spectral-normalized Gaussian process (SNGP) head~\cite{liu2020simple} is added to the model for the CTR prediction and is trained jointly with the main CTR head, producing a per-item posterior variance estimate. At serving time, a boost proportional to this uncertainty is multiplied to the main model's score~\cite{auer2002finite, guo2020deep}. The model is otherwise unchanged. Hypothesis: Explicitly shifting the ranking policy to increase exposure for uncertain content will help the model understand new items faster. In our experiments we also share the results from adding the head without using the prediction in any downstream ranking layer (5 (a). in Table \ref{tab:results}).\\

\noindent\textbf{Approach 6 -- Uncertainty-weighted training loss.} The same SNGP head supplies uncertainty estimates, but instead of being combined with the serving score, they are used to up-weight high-uncertainty training examples in the loss for the CTR head, in the spirit of curriculum learning~\cite{bengio2009curriculum} and self-paced approaches that emphasize examples where the model has the most to learn. The serving stack is unchanged. Hypothesis: the model would have low confidence on fresh or novel content predictions, and focusing learning on these items would improve model's understanding on these slices.

\section{Experimental Setup and Results}\label{sec:setup}
We report online A/B results for six singleton interventions and one combination experiment in \autoref{tab:results}, evaluated against the same control on the YouTube Music homepage over a two-week window affecting millions of daily visitors for each arm. Columns capture the metrics we use to argue about new-release surfacing (1-day and 7-day new-release engagement), discovery (novel-music engagement at the track level), surface composition (artist diversity, homepage repetition), and overall homepage engagement. We intentionally focus on engagement and exposure shifts rather than slice-specific ranking metrics such as AUC. An intervention might increase exposure simply by overpredicting scores for new items rather than ranking them more accurately. However, breaking a feedback loop inherently requires trading short-term precision for exploration. The immediate goal is to generate the exposure and interactions necessary to create training data, which has a second-order effect to improve the model's accuracy on new releases over a longer horizon. Rows correspond to the interventions, grouped by the layer of the system at which each is applied; the final row is a combination of the bias tower (layer 3) and the uncertainty-weighted loss (layer 4) discussed in \autoref{sec:methods}. 

The main observation is that no singleton intervention dominates across columns. Each row trades engagement, diversity, repetition, and new-release lift against one another, and the pattern of trades aligns with the layer at which the intervention is applied. Three patterns emerge among the singletons, and a fourth emerges from the combination experiment.

A separate observation we cannot fully explain is that simply adding the SNGP uncertainty head to the model and \emph{not} using its output at serving still produces a $+5.32\%$ 1-day new-release engagement lift with no significant engagement, novelty, or diversity cost. We treat this as an open observation rather than a recommended intervention. There are three plausible mechanisms that could cause it: loss rebalancing across the multi-task objectives, generic auxiliary-task regularization, and uncertainty-target-specific representation effects in the shared layers. We plan to run future experiments disambiguating between these.

\section{Discussion}\label{sec:results}

\subsection{Serving-Layer heuristics get neutralized}

We tested two serving-layer interventions: a recency boost that adds a multiplicative bonus to items released within 7 days, and a diversification heuristic that spreads out items by days since release. Neither modifies the model or the training data; both operate on the score returned by the production ranker.
 
\begin{figure}[ht]
  \centering
  \includegraphics[width=0.9\linewidth ]{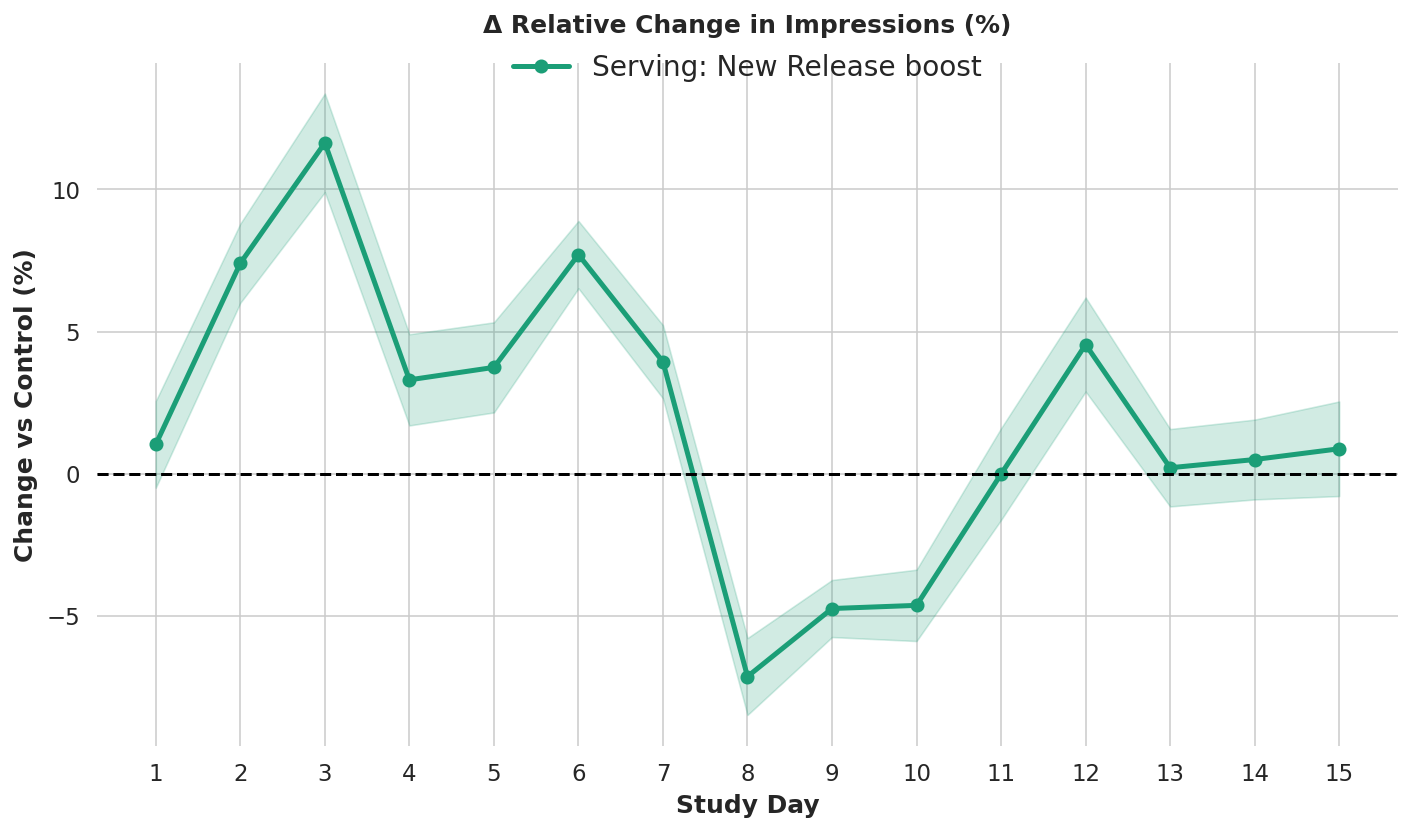}
  \caption{Change in impressions for New releases across the days of the heuristic recency boost at serving study}
  \label{fig:nr-boost-impressions}
  \Description{Diagram of Change in impressions for New releases across the days of the study}
\end{figure}

The recency boost shows no significant effect on either 1-day or 7-day new-release engagement. \autoref{fig:nr-boost-impressions} makes the underlying dynamic visible. Each eligible new release rendered on-screen to a user is counted as an impression. The daily lift in new-release impressions is positive and large in the first days of the experiment; other ranking components then react to the increase in unclicked impressions, drop the impressions sharply, and the metric bounces back to neutral within a few days. The boosted impressions enter the next training run, the model learns that the boosted items receive disproportionate exposure relative to their realized engagement, and the learned ranking partially undoes the boost. The YouTube Music homepage also has specific impression-based demotions that correct for overexposure faster than the ranker can adapt.

The general principle: any serving-time rule that the rest of the system can observe through its training inputs is subject to neutralization, and the more components of the stack are continuously trained, the larger the adaptation surface that can move against the rule. Serving-time interventions produce day-1 effects, but the effects are not durable, and the size of the eventual decay is not predictable from the day-1 measurement.

\subsection{Architectural debiasing reduces bias but does not improve freshness}\label{subsec:debias_freshness}

\begin{figure}[hbtp]
  \centering
  \includegraphics[width=1\linewidth ]{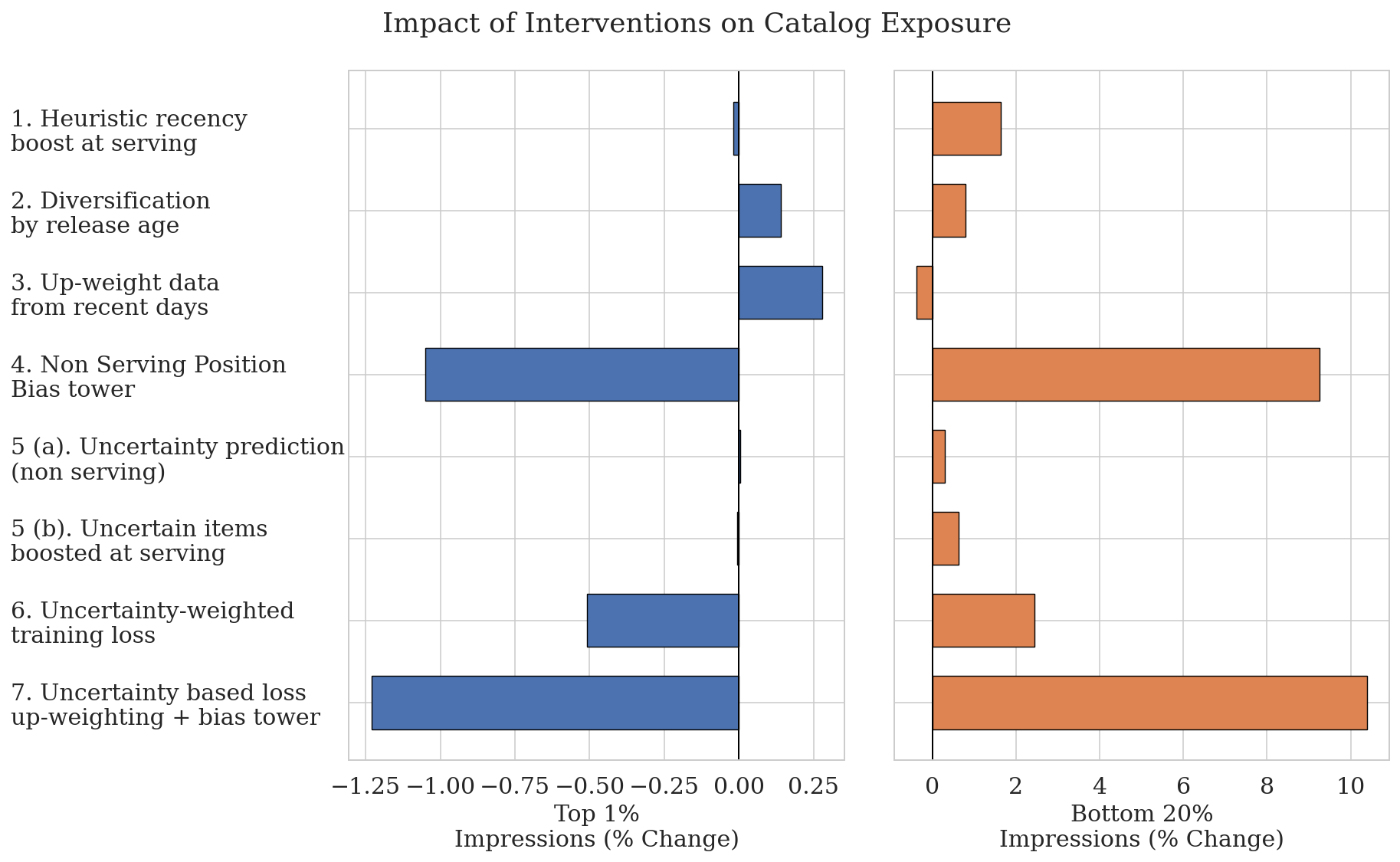}
  \caption{Change in impressions (\%) for the top and bottom impressed items across the methods. Top and bottom are determined by the total number of impressions for an item over the 2 week study period.} \label{fig:popularity_bias_fig}
  \Description{Diagram of the change in impressions for the top and bottom impressed items across the methods.}
\end{figure}

The bias tower produces the largest popularity redistribution of any singleton intervention we tested. \autoref{fig:popularity_bias_fig} shows the change in impression share for the top 1\% of items by historical popularity and for the bottom 20\% across all the methods; the bias tower delivers the largest reduction in top-1\% concentration and the largest expansion of bottom-20\% impressions ($+9.2\%$ relative to control), substantially larger than any other singleton. The  table corroborates the popularity-redistribution story along multiple metrics. Novel-music engagement increases by $4.70\%$ and Artist diversity increases by $0.79\%$ (the only singleton with a significant positive effect on diversity).

The bias tower is, however, not a new-release intervention. 7-day new-release engagement \emph{drops} by $3.39\%$ (CI $[-4.53, -2.24]\%$) and 1-day new-release engagement is flat ($+0.37\%$, not significant). New releases by popular artists were beneficiaries of the same popularity bias the tower removes, and the redistribution toward the unpopular tail does not propagate to new releases because new releases lack the historical engagement signal that would let the ranker learn to surface them. The bias tower removes a systematic distortion in the score; it does not add the new-content signal that surfacing a new release requires.

 \begin{figure}[h]
  \centering
  \includegraphics[width=\linewidth ]{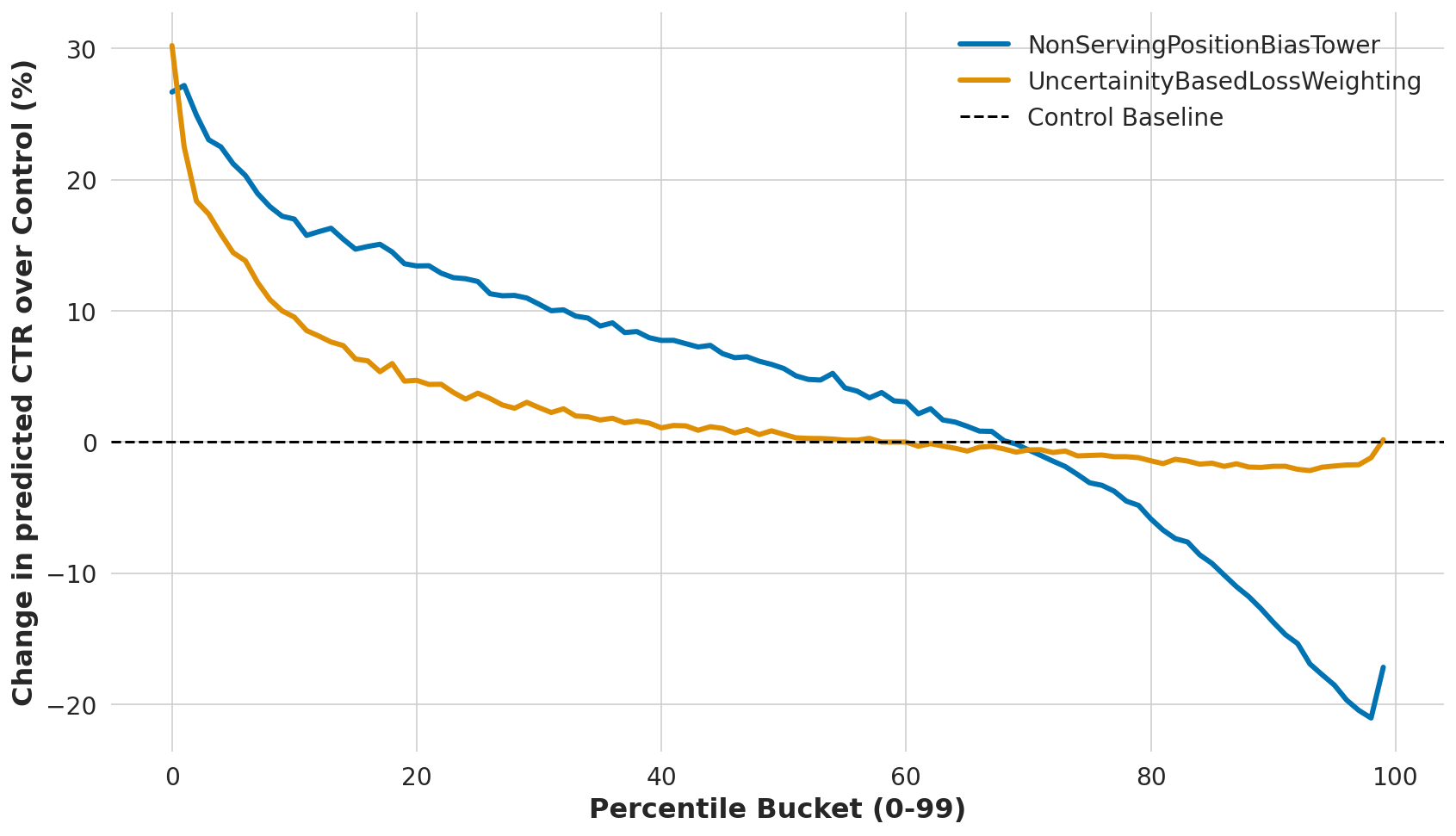}
  \caption{Relative change in predicted CTR per item, bucketed by control arm PCTR-percentile. The non-serving bias tower shifts the full distribution while the uncertainty-weighted training loss lifts only the bottom decile sharply with the upper end of the distribution minimally affected.} \label{fig:ctr_change_fig}
  \Description{Diagram of the change in impressions for the top and bottom impressed items across the methods.}
\end{figure}

One practical observation is that the bias tower shifts the predicted-score distribution. The CTR distribution flattens, with mass moving from the engaged head toward the middle (Figure \ref{fig:ctr_change_fig}). This is theoretically expected, since we drop the additive bias term during serving, but downstream consumers of the ranker's scores in our production stack (ranking functions, diversification, shelf rankers) are calibrated against the prior score distribution and require retuning. This cost is not unique to our system and practitioners considering a bias-tower-style intervention should budget engineering time for the recalibration.

\subsection{Uncertainty based exploration interventions produce the largest new release lift}

\begin{figure}[ht]
  \centering
  \includegraphics[width=0.9\linewidth ]{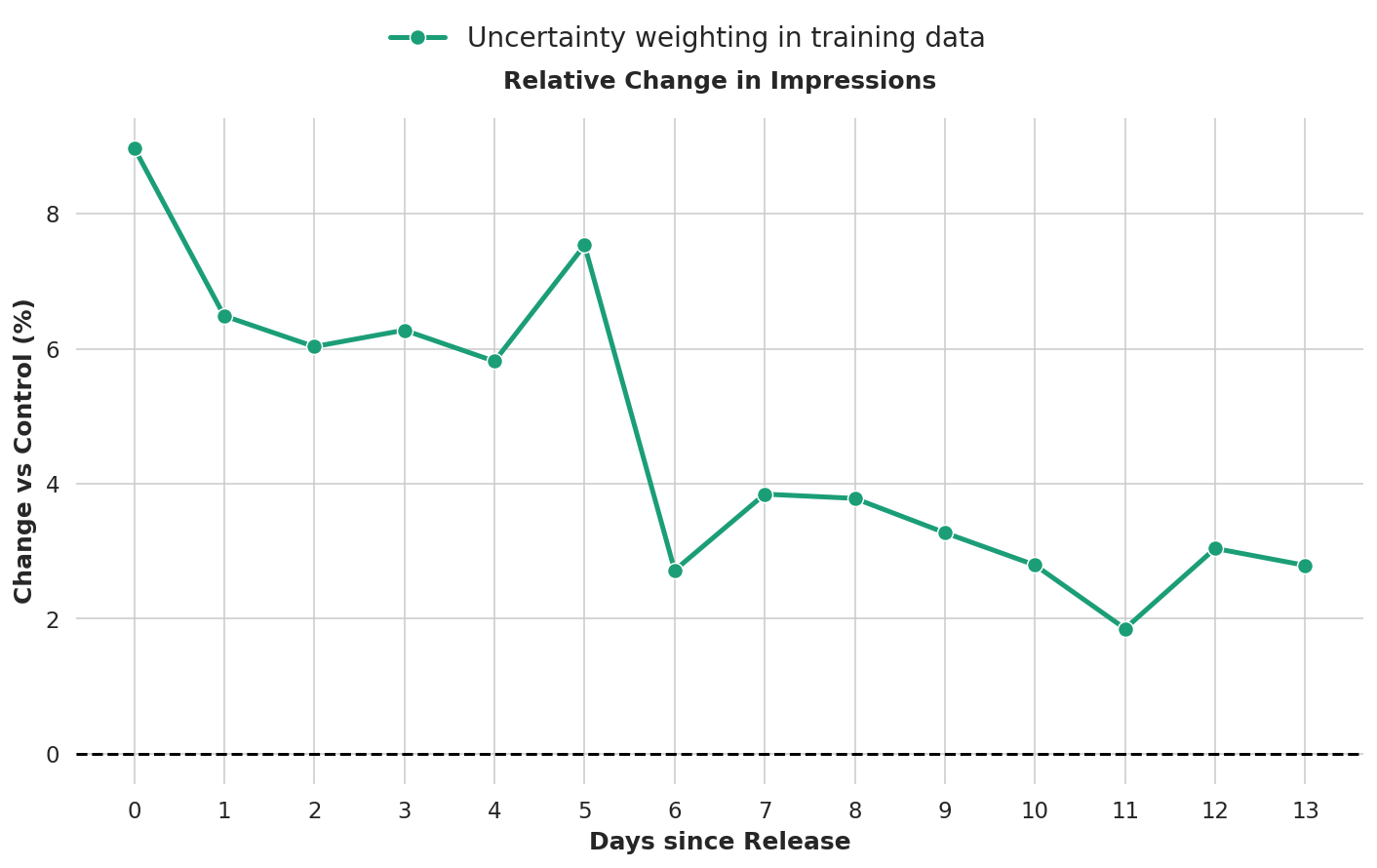}
  \caption{Change in impressions for New releases by the days since release of the item} \label{fig:uncertainity_weighting_impressions_change}
  \Description{Diagram of Change in impressions for New releases across the days of the study}
\end{figure}

Two interventions in our study use the SNGP-derived per-item uncertainty estimate to direct ranking attention toward items the model is least sure about. Boosting uncertain items at serving (5 (b). in Table \ref{tab:results}) adds a bonus to served scores at serving time. The uncertainty-weighted loss uses the same uncertainty signal to up-weight examples during training. Both produce the largest 1-day new-release engagement gains in the suite; they differ in cost structure and in how the gain is distributed across item age.

The uncertainty-prediction boost lifts 1-day and 7-day new-release engagement by the largest values in the new-release columns without negatively impacting Home engagement. But it drops Artist diversity and has only a minor impact on popularity bias in the system. The uncertainty-weighted loss on the other hand lifts 1-day new-release engagement but the change in 7-day new-release engagement is not significant. \autoref{fig:uncertainity_weighting_impressions_change} explains the gap. The lift is concentrated on the day-0 cohort and decays with item age. Each day as the model has learned enough about items in the dataset, the loss shifts to focus on new day-0 items. The loss up-weighting also drops home engagement and engagement on novel items. Uncertainty estimates from the SNGP head are concentrated on items lacking interaction history, and items lacking interaction history are heavily weighted toward new releases rather than toward the tail content that this specific user hasn't interacted with before.

This is direct empirical support for the freshness-versus-debiasing distinction set up in \autoref{sec:desc}. Even an intervention specifically designed to push exploratory content can prefer one kind of unfamiliar content over another in a way that changes the product outcome. 

\subsection{Composing layers: bias tower stacked with uncertainty-weighted loss}
The layered framework suggests that interventions in different layers should compose, with effects roughly additive excluding the second-order interactions through the shared training distribution. We tested this prediction directly by stacking the non-serving position bias tower with the uncertainty-weighted loss. The uncertainty prediction in this case was estimated for the \textit{unbiased} component of the CTR head which could also impact the distribution of the uncertainty predictions itself in addition to shifting the distribution of CTR. Because of production constraints, this arm ran over a one-week window rather than the two weeks that the other arms ran for. The cost of the shorter window is wider confidence intervals, which we flag where statistical significance is the load-bearing claim. The result is the final row of \autoref{tab:results}.

The combination does not cleanly add across all metrics :

\begin{itemize}
    \item The combination delivers a larger reduction in top-1\% impression share and a larger expansion of bottom-20\% impressions than the bias tower on its own (\autoref{fig:popularity_bias_fig}, last bar). Artist diversity is also slightly higher under the combination ($+0.90\%$) than under the bias tower alone. The bias tower removes the popularity-bias score component, while the uncertainty-weighted loss further reweights training toward low-history items, doubling down on the same redistribution.
    \item The 1-day new-release engagement lift, which the uncertainty-weighted loss singleton drives to $+4.33\%$ (significant, two-week window), drops to a point estimate of approximately $+2\%$ under the combination and is not statistically significant. The 7-day new-release engagement drop from the bias tower is nearly unchanged in the combination. 
    \item Homepage engagement drops by -0.43\% in the combination, close to the sum of the singleton effects. Homepage repetition falls by $0.48\%$, between the two singleton values. Novel-music engagement is $+1.85\%$, less than the singleton bias tower's and consistent with the negative novel-music effect of the uncertainty-weighted loss partially offsetting the bias tower's gain.
\end{itemize}

\subsection{Limitations}
The results reported here come from a single product (\ytm), a single surface (homepage), and a two-week experimental window per arm. Two weeks is enough to observe the temporal-decay dynamics that drive the serving-layer claim in Section 6.1, but it is short relative to the timescale on which long-run discovery effects on user retention or repertoire growth would materialize. Some of the new-release gains we report could be partially offset, or amplified, by longer-horizon dynamics that two weeks of online traffic cannot reveal. The combination experiment in Section 6.4 ran for one week rather than two, owing to production constraints. The resulting wider confidence intervals matter most for the 1-day new-release lift, where the two-week singleton significance does not transfer to the one-week combination.

We test one instantiation of each layer and only on the ranker. The bias-tower architecture follows WatchNext; another debiasing architecture such as a deeper bias network, or one that conditions on additional context, could trade differently between popularity redistribution and freshness. We did not run IPS as a live arm because of training instability in our setup, so the training-data layer is represented in our results only by recency upweighting. We do not claim the layered framework is a taxonomy of all possible interventions; we claim it is a useful lens for predicting where in the stack a given intervention's effects will and will not propagate.

\section{Conclusion}
We presented online A/B results for six interventions across four layers of a production music recommender, plus one combination arm, all evaluated against the same control on the YouTube Music homepage. The interventions are organized by where in the production stack they apply, and the overall finding is that the layer matters: which symptom of the feedback loop an intervention addresses, and which costs it incurs, is largely determined by its layer of application.

Serving-time heuristics produce real day-1 effects but are neutralized by the learning loop over the following days, because their outputs become training inputs the next time the model retrains. Architectural debiasing with a bias tower redistributes impressions away from the popular head and toward the long tail, and significantly improves engagement on novel items. Uncertainty-driven exploration with an SNGP head produces the largest sustained new-release lift; the gain is concentrated on day-0 items when the uncertainty signal is applied as a training-loss reweighting and is spread more broadly across the 7-day window when applied as a serving-time boost. Stacking the bias tower with the uncertainty-weighted loss composes super-additively on popularity redistribution and artist diversity but negates the new-release lift seen in the individual arm.  For practitioners, the choice of layer is itself a design choice with downstream consequences for the rest of the production stack. A practitioner choosing among these interventions should pick on the basis of which symptom they care most about — popularity dominance, freshness, or discovery — and budget for the integration cost of the layer where the intervention lives.

\begin{acks}
We would like to thank Yuening Li and Jiawei Li for their valuable technical consultations and feedback on integrating SNGP into the YouTube Music Ranking Model.
\end{acks}

\balance
\bibliographystyle{ACM-Reference-Format}
\bibliography{main}

\end{document}